\documentclass[12pt, draftclsnofoot, journal, letterpaper, onecolumn]{IEEEtran}

\IEEEoverridecommandlockouts
\usepackage{amsfonts}
\usepackage{graphicx}
\usepackage{times}
\usepackage{cite}
\usepackage{amsmath}
\usepackage{cases}
\usepackage{array}
\usepackage{dsfont}
\usepackage{verbatim}
\usepackage{amssymb}

\usepackage{stfloats}
\usepackage{authblk}
\usepackage{graphicx}
\usepackage{footnote}
\usepackage{booktabs}
\usepackage{array}
\usepackage{multirow}
\usepackage{bm}
\usepackage{empheq}
\usepackage[labelformat=simple]{subcaption}

\usepackage{amsthm}
\usepackage{algorithm}
\usepackage{algorithmic}
\usepackage{capt-of}

\usepackage{color}
\theoremstyle{plain}

\theoremstyle{definition}

\newtheorem{remark}{Remark}
\usepackage{multicol}
\usepackage{ dsfont }
\usepackage{multirow}
\newcommand{\at}[2][]{#1|_{#2}}

\begin{document}

\title{Management and Orchestration \\ of Virtual Network Functions \\ via Deep Reinforcement Learning}

\author[1]{Joan S. Pujol Roig}
\author[2]{David M. Gutierrez-Estevez}
\author[1]{Deniz G\"{u}nd\"{u}z}
\affil[1]{Imperial College London, London SW7 2AZ, UK}
\affil[2]{Samsung Electronics R$\&$D Institute UK, Surrey, TW18 4QE, UK }
\maketitle

\begin{abstract}
Management and orchestration (MANO) of resources by virtual network functions (VNFs) represents one of the key challenges towards a fully virtualized network architecture as envisaged by 5G standards. Current threshold-based policies inefficiently over-provision network resources and under-utilize available hardware, incurring high cost for network operators, and consequently, the users. In this work, we present a MANO algorithm for VNFs allowing a central unit (CU) to learn to autonomously re-configure resources (processing power and storage), deploy new VNF instances, or offload them to the cloud, depending on the network conditions, available pool of resources, and the VNF requirements, with the goal of minimizing a cost function that takes into account the economical cost as well as latency and the quality-of-service (QoS) experienced by the users. First, we formulate the stochastic resource optimization problem as a parameterized action Markov decision process (PAMDP). Then, we propose a solution based on deep reinforcement learning (DRL). More precisely, we present a novel RL approach called, parameterized action twin (PAT) deterministic policy gradient, which leverages an \textit{actor-critic architecture} to learn to provision resources to the VNFs in an online manner. Finally, we present numerical performance results, and map them to 5G key performance indicators (KPIs). To the best of our knowledge, this is the first work that considers DRL for MANO of VNFs' physical resources.
 
\end{abstract}

\section{Introduction}
Traditionally the deployment of new network functions (NFs) has been done through the acquisition and installation of a proprietary hardware running a licensed software. This fact reduces the incentives for network operators in updating their network's physical architecture to offer new services or update existing ones, as it represents an increase in both the capital expenditures (CAPEX), i.e., equipment inversion, equipment installation and personnel training, and operational expenditures (OPEX), i.e., the cost of of operating the system \cite{liang2015wireless}. To overcome this limitation, network function virtualization (NFV) has been proposed to curtail constant acquisition of technical hardware, by leveraging virtualization technology to implement NFs using general purpose computers/servers \cite{han2015network}. With virtualization, software implementation of a NF can be decoupled from the underlying hardware, i.e., NFs can be instantiated without the need of new equipment acquisition and installation, and they can run over commercial off-the-shelf hardware. The isolation of software from hardware allows for a set of VNFs to be deployed on a shared pool of resources. This motivates a solution to manage the underlying shared infrastructure (processing power, storage, etc.) in an efficient, scalable and rapid manner. 
\par 
There has been a lot of work on resource allocation for cloud networks. One of the most popular ways to address resource provisioning is threshold-based reactive approaches, where resources are added or removed if the network's condition reaches certain predefined thresholds \cite{murthy2014threshold,lorido2014review, aws, azure}. Although this provides a simple and scalable solution to dynamic resource allocation, threshold-based criteria tend to over-provision and under-utilize network equipment (incurring high costs for the infrastructure provider) and make the management of dynamic traffic and deployment of new types of services difficult as network traffic models must be elaborated beforehand.  In \cite{dutta2012smartscale}, authors study the scaling of virtual machines (VMs) in a proactive way. In particular they propose a solution via decision tree approach to resolve whether a VM instance should be \textit{vertically scaled}; that is, more physical resources (e.g., processing power, storage) should be added, or \textit{horizontally scaled}, i.e., by deploying a new VM instance. An autonomous vertical scaling approach is proposed in \cite{yazdanov2013vscaler} using Q-learning, where an agent learns how to autonomously provision resources (storage and processing power) to a VM.
\par 
With the explosion of machine learning (ML) and virtualization technologies, and their applications to communication networks, the idea of self-governing networks leveraging modern ML techniques is becoming popular among the communications research community. Chen et. al. \cite{chen2018optimized} proposes  deep double Q-learning (DDQ) and deep-SARSA solutions for mobile edge computing, where an end user terminal with limited local computation and energy resources jointly optimizes computation offloading and energy consumption selection in an autonomous manner. The end user terminal decides whether to execute a computing task locally or offload it to one or more of the available edge base stations (BSs), also selecting the amount of energy to be allocated for the task in question. A proactive VM orchestration solution is proposed in \cite{tang2015efficient} using Q-learning, where, given the current state, an agent decides to increase, reduce or retain the number of VMs allocated to a VNF. In \cite{rahman2018auto}, a deep learning approach is introduced to decide the number of VNFs that must be deployed to meet the network traffic demands. The authors formulate a classification problem, where each class corresponds to the number of VNFs that must be instantiated to be able to cope with the current traffic, and use historical labelled traffic data to train the proposed algorithm. 
\par 
More recently, in the management and  network orchestration (MANO) domain for wireless networks resources, the use of DRL has gained attraction for network slicing resource orchestration and management. These works formulate a discrete action selection optimization problem, and use well established value-based methods, e.g., Q-learning or SARSA, to solve the formulated problem. In this line lies the work of \cite{ zhao2018deep}, which proposes a deep Q-learning approach to radio resource slicing and priority-based core network slicing, showing its advantage in addressing demand-aware resource allocation. In  \cite{li2018deep}, authors formulate the problem of frequency bands allocation, and the problem of computation resources orchestration for different slices. These problems are reduced to choosing a particular configuration from a finite set of available configurations, which is done leveraging DDQNs. Similarly, in \cite{chen2019multi}, a DRL solution based on DDQN is presented for multi-tenant cross-slice resource orchestration, where again, a discrete number of communication and computation resources have to be allocated to different slice tenants. Finally a deep deterministic policy gradient (DDPG) with advantage function is employed in \cite{qi2019deep} to allocate bandwidth resources to different network slices. Compared to the aforementioned approaches, the continuous nature of DDPG allows for more fine-grained resource allocation. 
\par
In our work, we consider 3GPP functional split, where a central unit (CU) deploys and maintains a set of VNFs serving the users of several distributed units (DUs). We first formulate the dynamic allocation of processing and storage resources to VNFs as a Markov decision process (MDP). The optimal solution for this problem is elusive due to prohibitively large state and action spaces. Therefore, we present a novel deep reinforcement learning (DRL) algorithm, called  \textit{parameterized action twin} (PAT), where we use DDPG \cite{lillicrap2015continuous} and its novel variant called twin delayed DDPG  \cite{fujimoto2018addressing}, as well as ideas from the parameterized action Markov decision process (PAMDP) as in \cite{fujimoto2018addressing,hausknecht2015deep}, so that an agent placed at the CU is trained to learn whether to scale vertically (add processing power and storage), horizontally (instantiate new VNFs), or to offload (send the VNFs to the cloud) based on the system state (service request arrivals, service rates, service level agreement (SLA), etc.), using a cost function that combines the economic cost, SLA requirements, and the latency experienced by the users. The proposed algorithm is deployed in a variety of scenarios and its performance is evaluated according to a defined set of 5G key performance indicators (KPIs).
\par 
The feasibility of the proposed solution relies on the assumption that the technology envisaged for NF virtualization is \textit{``containerization"}\cite{sharma2016containers}, where  containers perform operating-system-level virtualization, i.e., every time a VNF is launched a container is deployed in a physical server. A ``container" \textit{is a lightweight, standalone, executable package of software that includes everything needed to run an application: code, runtime, system tools, system libraries and settings and its run by the operating system kernel} \cite{Docker}. Containers are isolated from one another and can communicate with each other through well-defined channels. We find containers to be a more appropriate  virtualization technology, compared to others, such as VMs, as they require less power \cite{morabito2015power}, take less start-up  and re-scaling time \cite{piraghaj2015framework,virtual}, and,  most importantly,  can be rescaled on-the-fly without disrupting the service they provide.
\par 
Due to the use of an actor-critic architecture, our approach is a joint policy and action-value-based optimization, which generally shows better convergence properties \cite{lillicrap2015continuous} compared to value-based approaches implemented in  \cite{tang2015efficient,chen2018optimized, zhao2018deep,  li2018deep, chen2019multi}. Moreover,  Q-learning and SARSA are used for discrete action selection \cite{sutton1998introduction}, which is not feasible for the continuous control problem addressed in this work. Although a continuous action space is considered in \cite{qi2019deep}, it focuses on the allocation of a single resource (bandwidth), while we consider the allocation of two continuous resources plus a discrete action for server selection. In contrast to \cite{tang2015efficient}, we consider not only horizontal scaling but also vertical scaling, as well as offloading, significantly increasing the complexity of the problem. Furthermore, our algorithm works in an online manner, i.e., dynamically adapting to the network traffic, which differs from \cite{rahman2018auto}, where the algorithms are trained using historic labelled data and cannot adapt to new types (i.e., classes) of traffic that differs significantly from the training set. Moreover, in disagreement with what is stated in \cite{rahman2018auto} for reinforcement learning (RL) approaches, our approach can use unlabelled historical data to learn, as we are interested in the network patterns (captured by the historical arrival and service times) to update the critic value-function estimates accordingly. Finally, in comparison with the cloud management algorithm presented in \cite{dutta2012smartscale}, using deep neural networks (DNNs) for function approximation can handle a higher dimensional state space, which would be challenging to be captured using decisions trees due to the exponential growth in the number of leaves.
\par 
The remainder of this paper is organized as follows: In Section \ref{sec: system model} the system model is introduced. The problem formulation using a Markov decision process (MDP) framework is presented in Section \ref{sec:Problem Formulation}. In Section \ref{sec:RL} we provide an overview of the RL notation, and  review the works upon which our approach is based. The proposed PAT algorithm used to train the agent is explained in Section \ref{s:PAT}. Numerical results illustrating the performance of the PAT algorithm are presented in Section \ref{s:NR}. Finally a summary of the results and conclusions is presented in Section \ref{s:conclusions}.
\par 
\underline{Notation:} $[\cdot]^{T}$ denotes the transpose operation. $\mathds{1}(x)$ denotes the logical operator, which equals to 1 if x is true, and 0 otherwise. For positive integer K, $[K]$ denotes the set $\{1,2,\dots, K\}$. $1_N$ denotes the vector of 1s of size N. For set $\mathcal{A}$, we denote its power set, i.e., the set of all subsets of $\mathcal{A}$ by $2^{\mathcal{A}}$. We define the function $clip\left(x,x_{min}, x_{max}\right) \triangleq \max \left\lbrace x_{min}, \min \left\lbrace  x_{max},x \right\rbrace \right\rbrace$.
\section{System Model}\label{sec: system model}

\begin{figure}[!h]
\centering
\includegraphics[width=0.7\textwidth]{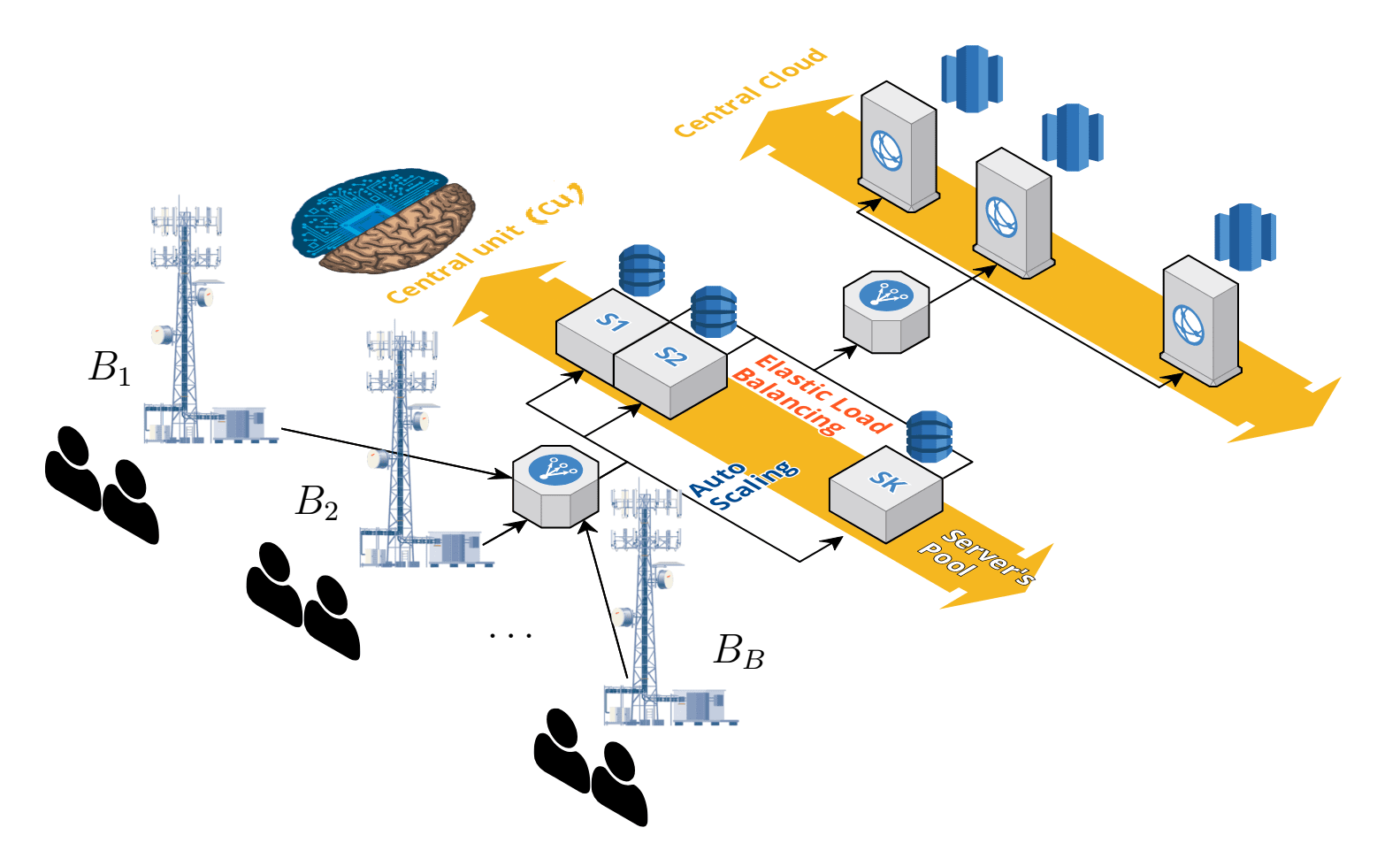}
\caption{Considered system architecture.}
\label{fig:mano}
\end{figure}

We consider a radio access network (RAN) with the 3GPP CU-DU functional split, consisting of $B$ small-cell BSs (the DUs),  denoted by $\mathcal{B} = \{\mathrm{B_1}, \mathrm{B_2}, \dots, \mathrm{B_B}\}$, connected to a CU that is in charge of the MANO of the NFs, such as transfer of user data, mobility control, RAN sharing, positioning, session management, etc. The BSs work as remote radio heads (RRHs), i.e., relaying all its traffic to the CU. Let $\mathcal{N} = \{\mathrm{N_0}, \dots, \mathrm{N_{N-1}}\}$ denote the set of distinct heterogeneous NFs offered by the CU that can be instantiated by any traffic requirement in the network. Based on the users' traffic requirements from  $B_i, i \in [B]$, the CU deploys and maintains a subset of VNFs from set $\mathcal{N}$. See Figure \ref{fig:mano} for an illustration of the network model.
\par
We envisage an autonomous CU that has a local pool of resources that  can be used to instantiate new VNFs, or to maintain deployed ones. The pool of local resources consists of $K$ servers denoted by $\{\mathrm{S_1},\dots, \mathrm{S_K}\}$, each with a limited processing and storage capability. We assume homogeneity across servers, such that each $S_k$, $ k \in [K]$, has the same storage size of $\eta _{max}F$ bits and a central processing unit (CPU) of capability $\rho _{max}C$ $\mathrm{Hz}$. In addition to the local servers, the CU can also employ resources located at a cloud center by offloading VNFs to the cloud, albeit at an increased cost which will be specified later.  We consider a central cloud with an infinite capacity resource pool, and denote it by $S_{K+1}$, so the set of available resources to the CU is denoted by  $\mathcal{K}=\{ \mathrm{S_1}, \dots, \mathrm{S_{K+1}}\}$.
\par
We consider a slotted resource allocation scheme, where new users arriving at the system wait until the start of the next slot to be allocated resources. Thus, the time horizon is discretized into decision epochs, corresponding to slots of duration $T$, and are indexed by an integer $t \in \mathbb{N}^+$.  At the beginning of each decision epoch, the CU decides how to allocate the network resources so that the VNFs can operate in the most agile and efficient way.  We denote by $\mathcal{N}^{\left(t\right)} \subset \mathcal{N}$, the set of active VNFs maintained by the CU during epoch $t$.  The dependency on $t$ of $\mathcal{N}^{\left(t\right)}$ emphasizes the fact that VNFs can be added or removed from the set of services provided by any BS overtime. Let us refer to $\rho _{k}^{\left(t\right)}\leq \rho _{max}$ and $\eta_{k}^{\left(t\right)}\leq \eta _{max}$ as the total processing power and storage, respectively, of server $S_k$ being used at epoch $t$. The CU is connected to the cloud via a dedicated link of capacity $R^{\left(t\right)}$ Mbps. We denote by $N_{k,j}$  the instance of $N_j$ deployed at server $S_k$, and allow only for one instance of each VNF to be deployed in a server.
\par 
We consider two main physical resources to be provisioned to the VNFs, CPU and memory, and assume that NF instance $N_{k,j}, \forall j \in [N]$, at epoch $t$ uses $c^{\left(t\right)}_{k,j}C\ \mathrm{H_z}$ of CPU capability and $m_{k,j}^{\left(t\right)}B$ bits of storage, at  server $S_k$, $k \in [K+1]$, where $c_{k,j}^{\left(t\right)} \in \left[\underline{c}_{k,j}^{\left(t\right)},\overline{c}_{k,j}^{\left(t\right)}\right]$ and $m_{k,j}^{\left(t\right)} \in \left[\underline{m}_{k,j}^{\left(t\right)},\overline{m}_{k,j}^{\left(t\right)}\right]$. Thus, each VNF has a different resource range in which it can operate following the definition of an \textit{elastic NF}, which refers to a VNF whose QoS \textit{gracefully} degrades with the scarcity of resources \cite{gutierrez2018path}. The range of CPU and memory resources VNF $N_{k,j}$ is able to operate at is given by:
\begin{equation}\label{eqn:cpu}
\begin{cases}
\underline{c}_{k,j}^{\left(t\right)}=c_{j,0}+\left(c_{j,r}- \Delta c_{j,d}\right)u_{k,j}^{\left(t\right)}\\
\overline{c}_{k,j}^{\left(t\right)}=c_{j,0}+\left(c_{j,r}+ \Delta c_{j,d}\right)u_{k,j}^{\left(t\right)}\\
\underline{m}_{k,j}^{\left(t\right)}=m_{j,0}+\left(m_{j,r}- \Delta m_{j,d}\right)u_{k,j}^{\left(t\right)}\\
\overline{m}_{k,j}^{\left(t\right)}=m_{j,0}+\left(m_{j,r}+ \Delta m_{j,d}\right)u_{k,j}^{\left(t\right)}\\
\end{cases},
\end{equation} 
where $u_{k,j}^{\left(t\right)}$ denotes the number of users being served by VNF instance $N_{k,j}$ at epoch $t$; $c_{j,0}$ and $m_{j,0}$ represent the offset CPU and memory requirements, respectively, that do not depend on the number of users being served. The variables $c_{j,r}$ and $m_{j,r}$ account for the linear increment of CPU and memory per user being served by the particular deployment of $N_j$ in server $S_k$. Values $\Delta c_{j,d}$ and $\Delta m_{j,d}$ are referred as the \textit{elastic service coefficients}, and define the resource range under which the VNF is able to operate. 
\par 
The QoS of the $u_{k,j}^{\left(t \right)}$ users served by the instance of VNF $N_{k,j}$  is denoted by $\mathrm{QoS}_{k,j}^{\left(t \right)}$, and depends on the resources allocated to this VNF instance. VNF $N_j$, $j \in [N]$, has a minimum QoS requirement, $\mathrm{QoS}_{j}^{min}$, that must always be ensured as specified by the SLA, and a maximum perceived QoS, $\mathrm{QoS}_{j}^{max}$. 

We assume that $\mathrm{QoS}_{k,j}^{\left(t \right)}$ as a function of $m_{k,j}^{\left(t \right)}$ and $c_{k,j}^{\left(t \right)}$ is given by the following piecewise function:
\begin{equation}\label{eqn:QoS}
\mathrm{QoS}_{k,j}^{\left(t \right)}=
 \begin{cases}
\mathrm{QoS}_j^{max}, \text{ if } c_{k,j}^{\left(t \right)} > \overline{c}_{k,j}^{\left(t\right)}\text{ and } m_{k,j}^{\left(t \right)} > \overline{m}_{k,j}^{\left(t\right)}\\
0,\text{ if } c_{k,j}^{\left(t \right)} < \underline{c}_{k,j}^{\left(t\right)}\text{ or } m_{k,j}^{\left(t \right)} < \underline{m}_{k,j}^{\left(t\right)}\\
\frac{\mathrm{QoS}_j^{max}-\mathrm{QoS}_j^{min}}{\overline{r}_{k,j}^{\left(t\right)}-\underline{r}_{k,j}^{\left(t\right)}}\left(\min\{m_{k,j}^{\left(t \right)},\overline{m}_{k,j}^{\left(t \right)}\}+\min \{c_{k,j}^{\left(t \right)},\overline{c}_{k,j}^{\left(t \right)}\}\right)+\frac{\mathrm{QoS}_j^{min}\overline{r}_{k,j}^{\left(t\right)}-\mathrm{QoS}_j^{max}\underline{r}_{k,j}^{\left(t\right)}}{\overline{r}_{k,j}^{\left(t\right)}-\underline{r}_{k,j}^{\left(t\right)}}, \\ \text{ otherwise}
\end{cases},
\end{equation}
where we defined $\overline{r}_{k,j}^{\left(t\right)} \triangleq \overline{c}_{k,j}^{\left(t\right)}+\overline{m}_{k,j}^{\left(t\right)}$ and $\underline{r}_{k,j}^{\left(t\right)} \triangleq \underline{c}_{k,j}^{\left(t\right)}+\underline{m}_{k,j}^{\left(t\right)}$. We see that $N_{k,j}$ satisfies the SLA if and only if  $m_{k,j}^{\left(t \right)}\geq \underline{m}_{k,j}^{\left(t\right)}$ and $c_{k,j}^{\left(t \right)}\geq \underline{c}_{k,j}^{\left(t\right)}$ as these result in $\mathrm{QoS}_{k,j}^{\left(t \right)}\geq \mathrm{QoS}_j^{min}$, and that additional resources  beyond  $\overline{c}_{k,j}^{\left(t\right)}$ and $\overline{m}_{k,j}^{\left(t\right)}$ do not have an impact on the QoS perceived by the users, which saturates at $\mathrm{QoS}_j^{max}$. Furthermore, Eqn. (\ref{eqn:QoS}) the fact that the QoS is the same for all the users being served by the same VNF  instance $N_{k,j}$, i.e.,  $u_{k,j}^{\left(t \right)}$. The CU can also offload VNFs to the cloud, in which case the CU's local pool are not used. 
\par 

We define $\textbf{c}^{\left(t\right)}_{k}\triangleq \left[c^{\left(t\right)}_{k,1}, \dots ,c^{\left(t\right)}_{k,N}\right]^{\intercal}$ and $\textbf{m}^{\left(t\right)}_{k}\triangleq\left[m^{\left(t\right)}_{k,1}, \dots ,m^{\left(t\right)}_{k,N}\right]^{\intercal}$. Finally, the matrices
$\textbf{C}^{\left(t\right)}\triangleq[\textbf{c}^{\left(t\right)}_{1}, \dots ,\textbf{c}^{\left(t\right)}_{K}]$ and 
$\textbf{S}^{\left(t\right)}=[\textbf{m}^{\left(t\right)}_{1}, \dots ,\textbf{m}^{\left(t\right)}_{K}]$ represent the overall CPU and memory allocations across all the K servers at epoch $t$, such that  $\rho _{k}^{\left(t\right)} = 1_{N}\cdot \textbf{c}_{k}^{\left(t\right)}$ and  $\eta _{k}^{\left(t\right)} = 1_{N }\cdot \textbf{m}_{j}^{\left(t\right)}$.
\par

The number of new service requests from all the BSs for VNF $N_j, j \in [N]$, in epoch $t$, is denoted by $n_{j}^{\left(t\right)}$, and is assumed to follow an independent and identically distributed (i.i.d.) homogeneous Poisson  process with parameter $\lambda _{j} ^{\left(t\right)}$; in other words, the probability of $n_{j}^{\left(t\right)}$ new demands to arrive at the CU for VNF $N_j$ in epoch $t$ for a time-slot of duration $T$ is given by:
\begin{equation}
P\left(n_{j}^{\left(t\right)}=n\right)= \frac{\left( \lambda _{j}^{ \left(t\right)}T\right)^n}{n!}\rm{e} ^{-\lambda _{j} ^{\left(t\right)}T}.
\end{equation}
\begin{remark}
In order to capture slow variations of network traffic over time, we consider time-varying $\lambda _{j}^{\left(t\right)}$ values, obtained by sampling a Gaussian distribution with parameters $\mu _j$ and $\sigma _j$ and taking the maximum between the obtained value and 0, i.e., $\lambda _{j}^{\left(t\right)}= \max \left\lbrace  x ,0 \right\rbrace$ where $x\sim\mathcal{N}\left( \mu _j,\sigma _j\right)$. We assume  value of $\lambda_j$ is kept constant for a block of $t_{max}$ time slots, and changes to an independent realization from the aforementioned truncated Gaussian distribution for the next block. 
\end{remark}
\begin{remark}
The value of $R^{\left(t\right)}$ is also obtained by sampling a truncated Gaussian distribution. Its values is given by $R^{\left(t\right)}=\max \left\lbrace  x ,R_{min} \right\rbrace$ where $x\sim\mathcal{N}\left( \mu _r,\sigma _r\right)$.
\end{remark}
We model users' service times by a geometric distribution; i.e.,  at the end of each time slot a user will remain in the system with probability $p _j$, and will leave the system with probability $1-p_j$, so the expected service time of a user is $1/p_j,\ \forall j \in [N]$. 
\par

There are three objectives the CU may want to optimize simultaneously: latency, financial cost and service quality. In order to simplify this multi-objective optimization problem we let the CU to minimize the long-term cost formed as a weighted combination of these three objectives. Next we explain each of these costs.
\subsubsection*{\textbf{Latency ($\delta _{T_{k,j}}^{\left(t\right)}$)} }
The latency cost associated with VNF instance $N_{k,j}$ during epoch $t$ is due to three potential causes of latency:
\begin{itemize}
\item \underline{VNF resizing latency $\left(\delta _{r_{k,j}}^{\left(t\right)}\right)$}:
This latency is associated with resizing the containers. Resizing a VNF consists of varying the amount of allocated CPU and memory resources. Docker allows to resize containers on-the-fly by using the command  \texttt{docker update} (from Docker v1.11.1). We consider that any instantiated container incurs  a delay of $\delta _{r,c}$ per unit $C$ of CPU added/removed, and $\delta _{r,m}$ per block of memory of size $F$ added/removed. Thus, the VNF instance $N_{k,j}$ resizing latency is given by:
\begin{equation}
\delta _{r_{k,j}}^{\left(t\right)}= |c_{k,j}^{\left(t\right)}-c_{k,j}^{\left(t-1\right)}|\delta _{r,c}+|m_{k,j}^{\left(t\right)}-m_{k,j}^{\left(t-1\right)}|\delta _{r,m}.
\end{equation}
\item \underline{Deployment latency $\left(\delta _{d_{k,j}}^{\left(t\right)}\right)$}: When a new VNF $N_{k,j}$ is instantiated  on a server $S_k, k \in [K]$, we consider a boot-up delay of $\delta _{d,b}$ per container. The total deployment latency of instance $N_{k,j}$ is given by 
\begin{equation}
\delta _{d_{k,j}}^{\left(t\right)} = \mathds{1}\left( c_{k,j}^{\left(t-1\right)}=0 \text{ and }  c_{k,j}^{\left(t\right)}>0\right)\delta _{d,b}.
\end{equation}
\item \underline{Offloading  latency $\left(\delta _{off_{K+1,j}}\right)$}: If a VNF instance is decided to be deployed on the cloud, in order to keep the service running, a continuous flow of information between the cloud and the CU is retained until the VNF is terminated.  This communication incurs a total latency of \begin{equation}
\delta _{off_{K+1,j}}=2 \overline{m}_{K+1,j}B/R^{\left(t\right)}\end{equation} for the offloaded VNF. Once a VNF is deployed in the cloud, we consider that the maximum resource utilization is guaranteed $\overline{r}_{j}^{\left(t\right)}$, so that $\mathrm{QoS}_{K+1,j}^{\left(t\right)}=\mathrm{QoS}^{max}$. 
\end{itemize}
The total latency incurred by VNF instance $N_{k,j}$ at epoch $t$ is given by:
\begin{equation} \label{eqn:latency_cost}
\delta _{T_{k,j}}^{\left(t\right)} = u_{k,j}^{\left(t\right)}\cdot \begin{cases}
\delta _{d_{k,j}}^{\left(t\right)}+\delta _{r_{k,j}}^{\left(t\right)}, \text{ if } k \in [K]\\
\delta _{off_{K+1,j}}, \text{ otherwise}
\end{cases}.
\end{equation}
All  the users, being served by instance $N_{k,j}$ experience the same latency, and hence the scaling by $u_{k,j}^{\left(t\right)}$, the number of users being served for each instance.  
\par

\subsubsection*{\textbf{Financial Cost ($C_{T_{k,j}}^{\left(t\right)}$)} } A price model that takes into account the economic implications of each $N_{k,j}$ VNF instance configuration is developed. 
\begin{itemize}
\item \underline{Resource cost  $\left(C_{r_{k,j}}^{\left(t\right)}\right)$:} We consider a financial cost of  $C_{r,m}$ per B bits of memory per epoch and $C_{r,p}$ per $C$ units of CPU resource per epoch for  server $S_k, k \in [K]$, i.e.,
\begin{equation}
C_{r_{k,j}}^{\left(t\right)}= c_{k,j}^{\left(t\right)}C_{r,p} + m_{k,j}^{\left(t\right)}C_{r,m}.
\end{equation}
\item \underline{Server cost $\left(C_{i_{k,j}}^{\left(t\right)}\right)$:} Every time a server $S_k, k \in [K]$, is powered on, we consider a one-time payment of $ C_{i,0}$ plus a rental cost of $ C_{i,v}$ per epoch. Hence the server cost is given by:
\begin{equation}
C_{i_{k,j}}^{\left(t\right)} = \mathds{1}\left( 1_{N}\textbf{c}_{k}^{\left(t-1\right)}=0 \text{ and }  1_{N}\textbf{c}_{k}^{\left(t\right)}>0\right)\frac{C_{i,0}}{N}+\mathds{1}\left(1_{N}\textbf{c}_{k}^{\left(t\right)}>0\right)\frac{C_{i,v}}{N}.
\end{equation}
\item \underline{Cloud cost$\left(C_{c_{K+1,j}}^{\left(t\right)}\right)$):} The financial cost of offloading a VNF to the cloud is modelled as a  one-time payment of $C_{c,0}$ plus a rental payment of $C_{c,v}$ per user per epoch until the VNF is terminated. Thus the cloud cost of VNF $N_j$ is given by:
\begin{equation}
C_{c_{K+1,j}}^{\left(t\right)} = \mathds{1}\left( c_{K+1,j}^{\left(t-1\right)}=0 \text{ and }  c_{K+1,j}^{\left(t\right)}>0\right)C_{c,0}+\overline{m}_{K+1,j}C_{c,v}.
\end{equation}
\end{itemize}
The total financial cost of VNF instance $N_{k,j}$ at epoch $t$ is given by:
\begin{equation}\label{eqn:financial_cost}
C_{r_{k,j}}^{\left(t\right)}= u_{k,j}^{\left(t\right)}\cdot\begin{cases}
 C_{r_{k,j}}^{\left(t\right)}+C_{i_k}^{\left(t\right)}, \text{ if } k \in [K]\\
C_{c_{K+1,j}}^{\left(t\right)}, \text{ otherwise}
\end{cases},
\end{equation}
\subsubsection*{\textbf{Service Level Agreement ($SLA_{k,j}^{\left(t\right)}$)} }
Each VNF instance $N_{k,j}$, $k \in [K]\ j \in [N]$, is associated with a minimum QoS requirement,  $\mathrm{QoS}_{j}^{min}$, and the failure to provision resources accordingly might incur service disruption, which violates the SLA. Accordingly, we define the SLA cost at VNF instance $N_{k,j}$ as
\begin{equation}\label{eqn:SLA}
SLA_{k,j}^{\left(t\right)}= \left(\gamma_{j} \mathds{1}\left( \mathrm{QoS}_{k,j}^{\left(t\right)}<\mathrm{QoS}_j^{min}\right)-\mathrm{QoS}_{k,j}^{\left(t\right)})\right)u_{k,j}^{\left(t\right)},
\end{equation}
where $\mathrm{QoS}_{k,j}^{\left(t\right)}$ is the perceived QoS of VNF $N_j$ at server $S_k$, and  $\gamma _{j}$ is the penalty for not fulfilling $\mathrm{QoS}_j$. Furthermore, the SLA cost scales with the number of users being served by the VNF instance $N_{k,j}$ as all of them experience the same QoS. As it is going to be detailed later, our models tries to minimize a cost function; and thus, we would like to reinforce the good actions that lead to higher QoS by reducing the cost or by even making it negative. To this end, we define the SLA cost to be inversely proportional to the perceived QoS. Furthermore, the whole of users served by the VNF instance experience the same SLA cost, that is why, this cost is again scaled by the number of users $u_{k,j}^{\left(t\right)}$.
\par 
We remark that, in order to capture the impact of reconfiguration of a VNF container on the whole network, each objective cost function is scaled by the number of users, such that a reconfiguration that affects more users is penalized/rewarded more than those affecting less users.

\subsubsection*{\textbf{Network Cost}}
 We define the overall network cost as the total cost incurred by all the instances deployed in the network at decision epoch $t$, defined as:
\begin{equation}\label{eqn:total cost}
C_{T}^{\left(t\right)}= \frac{  \sum _{j\in \mathcal{N}^{\left(t\right)}}\sum_{k \in [K+1]}  \omega_1\delta _{T_{k,j}^{\left(t\right)}}+\omega _3SLA_{k,j}^{\left(t\right)}+\omega _2C _{T_{k,j}}^{\left(t\right)}}{  \sum _{j\in \mathcal{N}^{\left(t\right)}}\sum_{k \in [K+1]}u_{k,j}},
\end{equation}
where the weights $\omega_1, \omega _2, \omega _3 \in \mathbb{R}^+$ are fixed weights independent of the VNF and the server. These weights can be tuned based on the preferences of the network operator, e.g., a network operator might be more concerned about reducing the economic cost rather than providing a high quality service, etc. The normalization by the number of users is to balance the network cost between heavy and low traffic periods. Without such a normalization busy traffic periods would incur higher costs regardless of the CU's performance. 

\subsubsection*{\textbf{VNF Instance Cost}} 
For purposes that will be explained in Section \ref{sec:RL}, we define the VNF instance cost $C_{k,j}^{\left(t\right)}$  incurred by instance $N_{k,j}$ at epoch $t$, as follows:

\begin{equation}\label{eqn:vnf cost}
C_{k,j}^{\left(t\right)}= \frac{ \omega_1\delta _{T_{k,j}^{\left(t\right)}}+\omega _3SLA_{k,j}^{\left(t\right)}+\omega _2C _{T_{k,j}}^{\left(t\right)}}{u_{k,j}}.
\end{equation}
This cost measures the contribution of a particular VNF instance to the global network cost. 
\section{Problem Formulation}\label{sec:Problem Formulation}
In this section, we formulate the  resource allocation problem as a MDP. We envisage an autonomous CU with the goal of  minimizing the long-term cost. To this end, we define the state space and the set of actions that the CU can take at each decision epoch.

\subsection{MDP}\label{ss:MDP}
At each decision epoch of a MDP an agent observes a state $s^{\left(t\right)}\in \mathcal{S}$, where $\mathcal{S}$ is the state space, and selects and action $a^{\left(t\right)} \in \mathcal{A}\left(s^{\left(t\right)}\right)$, where $\mathcal{A}\left(s^{\left(t\right)}\right)$ is the set of all possible actions in state $s^{\left(t\right)}$. Set $A =\cup_{s^{\left(t\right) }\in \mathcal{S}}\mathcal{A}\left(s^{\left(t\right)}\right)$ is referred as the action space. Action $a^{\left(t\right)}$ in state $s^{\left(t\right)}$ incurs a certain cost $R\left(s^{\left(t\right)},a^{\left(t\right)}\right)$, where $R: \mathcal{S} \times \mathcal{A}\rightarrow \mathbb{R}$ denotes the cost function, and the agent transitions to a new state $s^{\left(t+1\right)} \in \mathcal{S}$ with probability $p\left(s^{\left(t+1\right)}\mid s^{\left(t\right)},a^{\left(t\right)}\right)\in \mathcal{P}$, where $\mathcal{P}: \mathcal{S} \times \mathcal{A} \times \mathcal{S}\rightarrow [0,1]$ is a probability kernel. At each interaction, the agent maps the observed state $s^{\left(t\right)}$ to a probability distribution over the action set $\mathcal{A}\left(s^{\left(t\right)}\right)$. This MDP model is thus characterized by the 4-tuple $\langle s,a,r,p\left(s'\mid s,a\right)\rangle$.  This mapping specifies the \textit{policy} of the agent, and is denoted by $\pi$. The probability of selecting action $a^{\left(t\right)}=a$ in state $s^{\left(t\right)}=s$ is given by $\pi\left(a\mid s\right)$.

The \textit{state-value function}   $V_{\pi}\left(s\right)$ for policy $\pi$ at state $s$ is defined as the expected discounted cost the agent would accumulate starting at state $s$ following policy $\pi$:
\begin{equation*}
V_{\pi}\left(s\right) \doteq \mathop{\mathbb{E_\pi}} \left[\sum ^{\infty} _{t=1} \gamma ^{\left(t-1\right)} R\left(s^{\left(t\right)},\pi \left( s^{\left(t\right)}\right)\right) \mid s^{\left(1\right)}=s\right].
\end{equation*}
where $0\leq\gamma\leq 1$ is the discount factor that determines how far into the future the CU ``looks", i.e., $\gamma =0$ corresponds to a ``myopic" CU, that focus only on its immediate cost, while $\gamma =1$ represents an CU concerned with the cost over the whole time horizon. The action-value function, also referred as Q-function, is defined as: 
\begin{equation*}
Q_{\pi}\left(s,a\right) \doteq \mathop{\mathbb{E_\pi}}\left[\sum ^{\infty} _{k=0} \gamma ^k R\left(s^{\left(t+k\right)}, \pi \left( s^{\left(t+k\right)}\right)\right)\mid s^{\left(t\right)}=s,a^{\left(t\right)}=a\right].
\end{equation*}
We define the optimal value function, $V^*(s)$, as the minimum  expected total discounted cost obtained starting in state $s$ and following the optimal policy: 
\begin{equation}
V^*(s) = \min _{\pi} \mathbb{E}_{\pi}\left[V_\pi \left( s\right)\right].
\end{equation}
The goal is to find a policy $\pi*$ whose value function is the same as the optimal value function $V_{\pi*}\left(s\right)=V^*$.
\par 

Next, we define the state and action spaces for our problem.
\subsection{State Space}\label{ss:space}
The network state space is the set of all possible configurations of the network. The state at epoch $t$ consists of:
\begin{enumerate}
\item the number of arrivals for each VNF  $\left\lbrace n_{j}^{ \left(t\right)}\right\rbrace_{ j \in \mathcal{N}}$.
\item deployed  VNFs $\mathcal{N}^{\left(t\right)}$.
\item number of users being served by VNF $N_j$ at each server, $u^{ \left(t\right)}_{k,j}$.
\item cloud link capacity, $R^{\left(t\right)}$.
\item CPU  resources allocated  to each VNF at each server, $\textbf{C}^{\left(t\right)}$.
\item memory resources allocated to each VNF at each server,  $\textbf{S}^{\left(t\right)}$.
\end{enumerate}

The network state at epoch $t$ is characterized by $s^{\left(t\right)}=\left(\underset{ j \in \mathcal{N}}{\bigcup _{}}n _{j}^{ \left(t\right)},\ \mathcal{N}^{\left(t\right)},\underset{k \in \mathcal{K}, j \in \mathcal{N}}{\bigcup} u_{k,j}^{ \left(t\right)}\right.,$
 $\left. \textbf{C}^{\left(t\right)},\textbf{S}^{\left(t\right)}, R^{\left(t\right)}\right)\in \mathcal{S}$, where 
\begin{equation}
\mathcal{S}=\left\lbrace \left( \mathbb{Z}^+\right)^{\mathcal{N}} \right\rbrace \times 2^{\mathcal{N}}\times\left\lbrace \left(\mathbb{Z}^+\right)^{  \mathcal{K} \times \mathcal{N}} \right\rbrace \times \left\lbrace \left\lbrace 0,\dots, \rho_{max}\right\rbrace^{\mathcal{N}\times \mathcal{K}} \right\rbrace \times \left\lbrace\left\lbrace 0,\dots, \eta_{max}\right\rbrace ^{\mathcal{K}} \right\rbrace  \times \mathbb{R}^+.
\end{equation}
\subsection{Action Space}\label{ss:actions}
We consider three distinct ways the CU can react to variations in the workload: \textit{vertical scaling}, \textit{horizontal scaling} and \textit{ offloading}. The CU actions are taken at the user level, that is, the CU selects an action for each user request arriving at the system. This allows the CU to allocate users requesting the same VNF to different servers. 
\par 

Following \cite{narasimhan2015language}, we employ a PAMDP formulation, where a discrete action set is defined as $\mathcal{A}_D=\left\lbrace a_1, a_2, \dots, a_D \right\rbrace$, and each  action $a \in \mathcal{A}_D$ is associated with $n_{a}$ continuous parameters $\left\lbrace p_1^{a},\dots, p_{n_{a}}^{a} \right\rbrace$, $p_i^a  \in \mathbb{R}$. Thus, each tuple $\left(a, p_1^{a},\dots, p_{n_{a}}^{a}\right)$ represents a distinct action, and the action space is given by $\mathcal{A}=\cup _{a \in \mathcal{A}_D}\left\lbrace a, p_1^{a},\dots, p_{n_{a}}^{a}\right\rbrace$. In our problem, the first discrete component denotes the server at which the user is assigned to, while the remaining continuous components denote how the associated resources are updated. 
\par

\subsubsection*{\textbf{Vertical Scaling}} The vertical scaling action space, denoted by $\mathcal{A}_{V}=[K]$, refers to actions adding resources to, or removing from, a deployed VNF container instance  at epoch $t$ \cite{dutta2012smartscale}.  Taking into account the traffic fluctuations and VNF requirements, a CU might decide to increase (decrease) the CPU, and/or memory resources allocated to a deployed VNF instance independently, i.e., the memory allocation can be increased while the CPU allocation is decreased, or vice-versa. Hence, we define the vertical scaling actions separately for the CPU and memory resources, as $p_{CPU}^{\left(t\right)}$ and $p_{M}^{\left(t\right)}$, respectively, as the change in the allocated resources with respect to time slot $t-1$. We have

\begin{equation}
\begin{cases}
p_{CPU}^{\left(t\right)}\in \left\lbrace i\cdot B\mid \ i \in \mathbb{R},\ -\rho_k^{\left(t\right)}\leq i \leq \rho_{max}-\rho_k^{\left(t\right)}\right\rbrace\\
p_{M}^{\left(t\right)}\in \left\lbrace i\cdot C\mid \ i \in \mathbb{R},\ -\eta_k^{\left(t\right)}\leq i \leq \eta_{max}- \eta_k^{\left(t\right)}\right\rbrace
\end{cases}.
\end{equation}
Vertical scaling is limited by the resources of the physical server in which a container is deployed, thus, the limitation of $ \rho_{max}-\rho_k^{\left(t\right)}$ and $\pm \eta_{max}-\eta_k^{\left(t\right)}$.
\par
Note that the parameters $p_{CPU}^{\left(t\right)}$ and $p_{M}^{\left(t\right)}$ represent  increment/decrement of the resources already allocated to VNF $N_j$ at server $S_k$, i.e., $c_{k,j}^{\left(t\right)}=c_{k,j}^{\left(t-1\right)}+p_{CPU}^{\left(t\right)}$ and  $s_{k,j}^{\left(t\right)}=s_{k,j}^{\left(t-1\right)}+p_{M}^{\left(t\right)}$; hence, $p_{CPU}^{\left(t\right)}$ and $p_{M}^{\left(t\right)}$ can also take negative values. As mentioned before, all the users of a server's VNF instance equally share the allocated resources, thus, all of them are affected by the reshuffling of resources.

\subsubsection*{\textbf{Horizontal Scaling}} 
Horizontal scaling refers to the deployment of new containers to support an existing VNF $N_j$ at epoch $t$. If the load of a VNF increases, and the CU estimates that server $k$ at epoch $t+1$ will not be able to support its operation the CU might create another instance of the same VNF in another server. \par 
We have  $\mathcal{A}_{H}=[K]$  and 
\begin{equation}
\begin{cases}
p_{CPU}^{\left(t\right)}\in \left\lbrace i\cdot B\mid \ i \in \mathbb{R},\ -\rho_k^{\left(t\right)}\leq i \leq \rho_{max}-\rho_k^{\left(t\right)}\right\rbrace\\
p_{M}^{\left(t\right)}\in \left\lbrace i\cdot C\mid \ i \in \mathbb{R},\ -\eta_k^{\left(t\right)}\leq i \leq \eta_{max}- \eta_k^{\left(t\right)}\right\rbrace
\end{cases},
\end{equation}
where $k$ denotes the server at which a new VNF instance is to be deployed using horizontal scaling.  $p_{CPU}^{\left(t\right)}$ and $p_{M}^{\left(t\right)}$ account for the amount of CPU and memory resources to be allocated for the new deployment of $N_{k,j}$ at server $k$.

\subsubsection*{\textbf{Work offloading}}
If the CU foresees that it cannot cope with a traffic fluctuation by scaling vertically or  horizontally,  it can decide to offload a VNF to the cloud. We define the offloading action as $\mathcal{A}_{off}^{\left(t\right)}$.
This action is not associated with any parameter due to the assumption of unlimited CPU and memory resources at the cloud.

\par
\subsubsection*{\textbf{Parametrized Action Space}}

Following the PAMDP notation the complete parameterized action space at epoch $t$ is given by
\begin{equation}
\mathcal{A}^{\left(t\right)}\triangleq\left(\mathcal{A}_{V}^{\left(t\right)}, p_{CPU}^{\left(t\right)},p_{M}^{\left(t\right)},\right)\cup\left(\mathcal{A}_{H}^{\left(t\right)},  p_{CPU}^{\left(t\right)},p_{M}^{\left(t\right)}\right)\cup \left\lbrace \mathcal{A}_{off}^{\left(t\right)}\right\rbrace,
\end{equation} 
so that the CU action taken at epoch $t$, $a^{\left(t\right)}\in \mathcal{A}^{\left(t\right)}$.\par 


The cost function for our problem has been defined in Section \ref{sec: system model} in detail. Note that the CU's action at each time slot consists of $n_j(t)$ actions in the PAMDP formulation, one for each user request. Note that the randomness in our problem is due to random users arrivals for each VNF, and the random service time for each user in the system. If these statistics are known, the optimal policy can be identified through dynamic programming (DP), e.g., by the value iteration algorithm. However, estimating these probabilities for our problem, which has large state and action spaces is prohibitive, making the DP solution practically infeasible. Hence, we will instead exploit DRL to find an approximation to the optimal value function.

\section{Reinforcement Learning For VNF Management}\label{sec:RL}

In the RL method used in this work, the agent does not necessarily exploit (or even know) the transition probabilities governing the underlying MDP as it learns directly a policy as well as the action-value functions based on its past experience (model-free). The formulated problem suffers from the curse of dimensionality due to the prohibitively large
size of the state and action spaces (continuous state space). Therefore, we employ the actor-critic method with NNs as a function
approximation to parametrize the policy, and allows the learning agent to directly search over the action space. Another DNN is employed to approximate the state-value functions, which are used as feedback to determine how good the current policy is.
\subsection{Deep Reinforcement Learning (DRL)}
The use of DNNs as general function approximators have been proven to work very well in a wide range of areas, such as  computer vision, speech recognition, natural language processing and recently wireless networks \cite{zhang2018deep}.  Traditional RL methods struggle to address real-world problems due to their high complexity. In these problems, high-dimensional state spaces need to be managed in order to obtain a model that can generalize past experiences to new states. For example, tabular Q-Learning uses a hash table to store the estimated cost of state-action pairs, so for continuous input states, even if quantizated,  this solution deems intractable, since even with modest 5-level quantization and a state vector of size N, $5^N$ entries would have to be stored ($\approx 10^{13}$ entries if $N=20$). DRL aims to solve this problem by employing NNs as function approximators to reduce the complexity of classical RL methods.
\par 
 In \cite{mnih2015human} authors introduce deep Q-learning network (DQN), where a DNN is used as a function approximator for action selection on a discrete action space, based on Q-Learning. Given a state, Q-Learning updates the action-value estimate with the immediate reward plus a weighted version of the highest Q-estimate for the next state. Using a combination of 3 convolutional layers (for computer vision) and two fully connected layers (Q-learning part), they obtain human-level results for a wide range of Atari games. Further architectures based on DQN have been proposed, such as Duelling DQN \cite{wang2015dueling}, where there are two distinct DNNs, one to estimate  $V_{\pi}\left(s\right)$ and the other to estimate the so-called \textit{advantage} function $A_{\pi}(s,a)=Q_{\pi}\left(s,a\right)-V_{\pi}\left(s\right)$. 
These methods work well for a continuous state space but are limited to discrete action space, suffering from the curse of dimensionality when the action space is large.
 \par
 To overcome the limitation of discrete action selection, in \cite{lillicrap2015continuous} the idea of DQN is extended to  continuous action spaces using the deterministic policy gradient (DPG) theorem \cite{silver2014deterministic}, in particular the deep-DPG (DDPG) method. DDPG extends the use of DNN to the actor-critic method leveraging off-policy learning, where a deterministic policy is learned using  a combination of \textit{replay buffer} and \textit{target networks} to ensure stability and a zero-mean Gaussian noise is added to the actions for action space exploration.
\par 
In \cite{fujimoto2018addressing} it is shown that DDPG may lead to overestimated action-value estimates, thereby to suboptimal policies. To overcome this problem, authors present a novel method called twin delayed DDPG (TD3). A novel actor-critic architecture is proposed which comprises two critic networks, hence, two different Q-functions are learned, and the smaller of the two estimates is used as the update rule for the critics. The proposed algorithm adds clipped noise to the target action to make it harder for the policy to exploit Q-function errors. They also propose to update the targets and the policy less frequently than the Q-functions, helping to reduce the variance.\par 
Given the high dimensionality of both the state and the action spaces in our model we propose a solution that leverages DNNs as policy and action-value function approximators, while exploiting the results from \cite{fujimoto2018addressing} for continuous action selection. To this end,  we implement a novel architecture for PAMDP to address the problem defined in Section \ref{sec:Problem Formulation}.
\section{Actor-Critic method in PAMDP}\label{s:PAT}
In the proposed approach, the actor-critic method is leveraged, which is a combination of value based and policy optimization approaches. It combines the benefits of both methods as the critic estimates the action-value function $Q_{\phi}(s,a)$, while the actor derives a policy $\pi_{\theta}(s)$  critically using the value estimates of the critic to update the policy. In this section we present our novel approach that implements the actor-critic method for a PAMDP, which we call the parameterized action twin (PAT). The proposed PAT algorithm is presented in Algorithm \ref{a:algorithm}.
For ease of notation in the rest of the section we will refer to $s^{\left(t\right)}=s$, $s^{\left(t+1\right)}=s'$, $a^{\left(t\right)}=a$, $R \left(s^{\left(t\right)},a^{\left(t\right)}\right)=r$.

\subsection{VNF MANO meets PAT}
Before detailing the proposed RL algorithm, we clarify its integration into the CU, and how it interacts with the environment described in Section  \ref{sec: system model}, as we believe it will ease the comprehension of the algorithm. 
\par
 At the beginning of each decision epoch $t$,  we randomly select a VNF and we proceed to serve its new arrived users. The random selection of VNF is motivated by fairness, so that we avoid starting the process of resource allocation (when more resources are available) with always the same VNFs. Following the random VNF selection, we iterate over all requests of this VNF to allocate the network resources using the PAT method. For resource allocation, a snapshot of the network state is used as input to the PAT method. Based on the network state, the proposed RL algorithm decides the actions to be taken and from which server/cloud the user is served. After the allocation, a new updated snapshot of the network state is obtained and the agent cost described in Section \ref{ss:agent reward} is computed. These transitions are stored in the memory buffer of the agent and will later be used to train the PAT algorithm, so that it adapts to previously seen as well as new traffic patterns. Even if a VNF does not have any new requests, we nevertheless select that VNF and apply the PAT algorithm for action selection so that, in case  it was already deployed, resources can be added or removed; if not, the VNF can be deployed ahead of future traffic. In this last case, the VNF instance only incurs an economical cost as the number of users being serve is 0.
\subsection{Parameterized Action Twin (PAT) algorithm}
Following \cite{fujimoto2018addressing}, we use two critics in order to obtain two distinct estimates of the action-values; thus, two different DNNs, parameterized by $\bm{\phi}_1 \text{ and }\bm{\phi}_2$, are used to estimate two different action-value functions.  The aim of the two critic networks, as explained in \cite{fujimoto2018addressing}, is to avoid overestimation. We find that clipping the critics' updates to the minimum between the two estimates yields better policies. Note that this update rule might introduce underestimation bias; however, we find it more convenient in the long term to avoid convergence to suboptimal policies. 
\par
\begin{figure}[!h]
\centering
\includegraphics[width=1\textwidth]{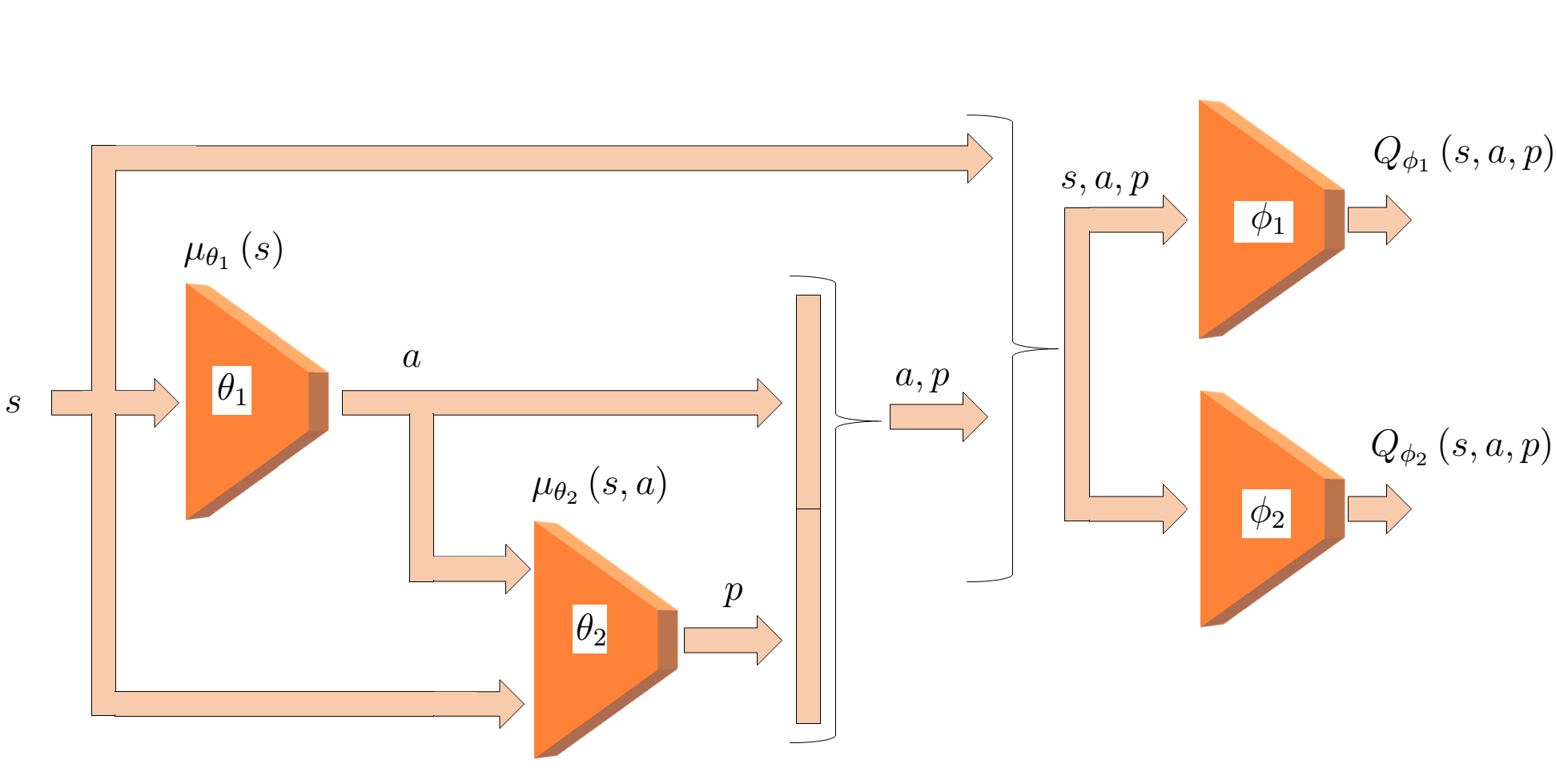}
\caption{The information flow between different DNNs in the proposed architecture.}
\label{fig:DNN_network}
\end{figure}

 
Two more DNNs, parameterized by $\bm{\theta}_1 \text{ and }\bm{\theta}_2$,  are used for the policy parameterization of the actor. The goal of the first actor network is to select the discrete action $a$ based on the current state $s$, while the second network generates the continuous action parameters $p=[p_{CPU}^{\left(t\right)},p_{M}^{\left(t\right)}]^T$ based on the outcome of the first actor network $a$ and the current system state $s$. Thus, the joint  selection $(a,p)$ determined by two distinct networks, in contrast to the approach in \cite{hausknecht2015deep}, where a single DNN architecture is used to determine both, the action and the parameters associated with it. We find this architecture to reflect a more natural process of action selection by first deciding the discrete action $a$, and then, choosing the associated parameters $p$ as defined in Subsection \ref{ss:actions}. We use a stochastic policy for discrete action selection while deterministic policy is leveraged for parameter selection, which we denote by $\mu$, i.e., parameters $\bm{\theta}_2$ map state and action $\left(s ,a\right)$ to parameters $\mu_{\theta_2}(s,a)=p$. Figure \ref{fig:DNN_network} illustrates the DNNs structure and the flow of information.
\par
Finally, four more DNNs are employed, corresponding to the mirroring target networks, and are parametrized by $\bm{\phi} _1^-,\bm{\phi} _2^-, \bm{\theta}_1^-, \bm{\theta}_2^-$, respectively. Their function is explained later in this section. 

\subsubsection{Parameter Updates}
 The critics take the network state $s$ and the action $(a,p)$, and estimate the value function $Q_{\phi_i}\left(s,a,p\right)$ , $i=1,2$. As is typical in actor-critic methods, we use off-policy temporal difference of 0, i.e., TD(0), for action-value function approximation, with clipped update rules as below:
\begin{equation}\label{eqn:q-learning}
Q^{t+1}_{\phi_i}\left(s,a,p\right)=Q^{t}_{\phi_i}\left(s,a,p\right)+ \alpha \left(r+\gamma \min _{i=1,2} \left\lbrace Q^{t}_{\phi_i^-}\left(s',\mu _{\theta^{-}_1} \left(s'\right),\mu _{\theta^{-}_2} \left(s',\mu _{\theta^{-}_1,} \left(s'\right)\right)\right)\right\rbrace-Q^{t}_{\phi_i}\left(s,a,p\right)\right),
\end{equation}
which minimizes the following loss function, for $i=1,2$:
\begin{equation}
L_{Q_{\phi_i}}\left(s,a,p\right)=\frac{1}{2}\left(r+\gamma \min _{i=1,2} \left\lbrace Q_{\phi_i^-}\left(s',\mu _{\theta^{-}_1} \left(s'\right),\mu _{\theta^{-}_2} \left(s',\mu _{\theta^{-}_1,} \left(s'\right)\right)\right)\right\rbrace-Q_{\phi_i}\left(s,a,p\right)\right)^2.
\end{equation}
The action-value functions of the critics are learned through gradient descent with the update rule:
\begin{equation}
\phi_i^{t+1}=\phi_i^{t}+\alpha \left(r+\gamma \min _{i=1,2} \left\lbrace Q_{\phi_i^-}\left(s',\mu _{\theta^{-}_1} \left(s'\right),\mu _{\theta^{-}_2} \left(s',\mu _{\theta^{-}_1,} \left(s'\right)\right)\right)\right\rbrace-Q_{\phi_i^{t}}\left(s,a,p\right) \right)\nabla_{\phi_i^{t}}Q_{\phi_i^{t}}\left(s,a,p\right).
\end{equation}
\par
The critics’ estimations of the joint action and parameters are gathered by the actors to update the policy. In continuous action space, the greedy update of the policy becomes infeasible as it requires a global maximization at every step, and going through all the action space to maximize  the estimated expected return is infeasible. Following\cite{silver2014deterministic}, we use the critics' network's gradient that indicates the direction the global Q-value estimate increases, to update the policy parameters. In order to obtain the gradients, we need to perform back-propagation through one of the critics’ network (we chose critic 1). It must be noted here that this gradient is not the conventional gradient over the network parametrization but with respect to the input, such that for the action network $\theta _1$ the update rule is 
\begin{equation}
\theta _1^{t+1}= \theta _1^{t} +\alpha \mathop{\mathbb{E}_{s \sim \rho_{\theta_1}}}\left[ \nabla_{\theta_1} \mu _{\theta_2}\left(s\right) \nabla_a Q_{\phi_1} \left(s,a,p \right)\at[\big]{a=\mu_{\theta_1}(s)}\right],
\end{equation}
while the update rule for network $\theta _2$  is
\begin{equation}
\theta _2^{t+1}= \theta _2^{t} +\alpha \mathop{\mathbb{E}_{s \sim \rho_{\theta_2}}}\left[ \nabla_{\theta_2} \mu _{\theta_2}\left(s,a\right) \nabla_p Q_{\phi_1} \left(s,a,p \right)\at[\big]{p=\mu_{\theta_2}(s,a)}\right],
\end{equation}
where $s \sim \rho_{\theta_i}$ refers to the trajectory sample using network $i$.
\par 
\begin{algorithm}[h]
\algsetup{linenosize=\small}
  \scriptsize

 \caption{Proposed PAT}
\begin{algorithmic}\label{a:algorithm}
\STATE Initialize the actors and critics networks, i.e., $\theta_1,\theta _2 \text{ and } \phi_1,\phi_2$ using Gaussian initialization with $\mu=0,\sigma = 10^{-2}.$
\STATE Copy the parameters to the target networks, i.e., $\phi _1^-\leftarrow \phi _1 ,\phi _2^-\leftarrow \phi _2, \theta_1^-\leftarrow \theta_1,\theta_2^-\leftarrow \theta_2 $
\STATE Initialize the replay buffer $\mathcal{M}$
\STATE $t=0$
\WHILE{$t<total\_timesteps$}
 \FOR{i= 0,...,T}
	\STATE Observe state $s$.
    \STATE Select action $a=\mu_{\theta_{2}}(s)$ with probability $1-\epsilon$ or a random action $a$ with probability $\epsilon$.
    \STATE Select parameter $p$ $$p=\text{clip}\left(\mu_{\theta_2}\left(s, a\right)+ w, p_{min}, p_{max}\right),$$ where $w \sim \text{clip} \left(\mathcal{N}\left( 0,\sigma^2\right)\cdot (p_{max}, \eta_{max}),-c,c\right)$
    \STATE Store transaction $\langle s,(a,p),r,s'\rangle \in \mathcal{M}$
    \ENDFOR
    \STATE Get a batch $\mathcal{B}=\{\left( s,(a,p),r, s'\right)\}$ of randomly sampled trajectories from the replay buffer $\mathcal{M}$
    \STATE Compute the action and parameter targets: 
    \begin{align*}
    a^-(s')&=\mu_{\theta_1^-}\left(s\right)\\
    p^-(s',a^-)&=\text{clip}\left(\mu_{\theta_2}\left(s', a^-(s')\right)+ w, p_{min}, p_{max}\right)
    \end{align*}
    \STATE Compute the targets estimates
    $$y(r,s')=r+\gamma \min _{i=1,2} \left\lbrace Q_{\phi_i^-}\left(s',a^- \left(s'\right),p^- \left(s',a^-(s')\right)\right)\right\rbrace$$
    \STATE Update the Q-Functions accordingly:
    $$ \nabla _{\phi_i}\frac{1}{|\mathcal{B}|}\sum _{\rho	\in \mathcal{B}}\left(Q_{\phi_i}\left(s,a,p\right)-y\left(r,s'\right)\right)^2\text{  for }i=1,2$$
    \STATE Update the action policy:
       $$ \nabla _{\theta_1}\frac{1}{|\mathcal{B}|}\sum _{(s,p)\in \mathcal{B}}Q_{\phi_1}\left(s,\mu_{\theta_1}(s),p\right)$$
    \STATE Update the parameter policy:
    $$ \nabla _{\theta_2}\frac{1}{|\mathcal{B}|}\sum _{(s,a)\in \mathcal{B}}Q_{\phi_1}\left(s,a,\mu_{\theta_2}(s,a)\right)$$
    \STATE Update the target networks:
    \begin{align*}
    \phi ^-_i&= \tau \phi_i + (1-\tau)\phi ^-_i\text{  for }i=1,2\\
\theta ^-&= \tau \theta_i + (1-\tau)\theta ^-_i\text{  for }i=1,2
\end{align*}
\STATE $t=t+1$
\ENDWHILE
\end{algorithmic}
\end{algorithm}
\subsubsection{Stabilizing updates}
Once both the critic and the actor networks are updated, the target networks should also be updated. Target networks are used to stabilize the updates. If the same $\phi_1=\phi_2=\phi$ network is used for bootstrapping ( i.e., estimating the value function of the next state $Q(s',a')$) and estimating $Q(s,a)$ in (\ref{eqn:q-learning}), the $\phi$ network will be updated with each iteration to move closer to the target Q-values; but, at the same time, the target Q-values, which are given by the same network, will also be changing in the same direction, like a dog chasing its tail. By introducing the target networks, we reduce this constant movement of the target estimates by delaying its update. The rule to update the target networks is given by
\begin{equation}
\begin{cases}
\phi ^-_i= \tau \phi^{t}_i + (1-\tau)\phi ^-_i\\
\theta_i ^-= \tau \theta^{t}_i + (1-\tau)\theta ^-_i
\end{cases},
\end{equation}
where $\tau\leq 1$ is an hyper-parameter used to regulate the speed at which the target networks are updated.
\par
Another tool to stabilize the network parameter updates is the memory buffer $\mathcal{M}$. The memory buffer stores the interactions of the agent with the environment, to be more precise, we store on-step trajectories, i.e., $s,a,r,s'$. Once the memory is filled with enough samples ($\approx 100K$), we uniformly sample the memory to obtain mini-batches of size $N=$ samples which are used to compute the losses of the actor and critic. The idea behind the use of a memory buffer is that, most optimization algorithms, including gradient descent, assume that the samples, from which the gradient estimate is obtained, are i.i.d. Clearly this is not the case in the defined environment; however, by sampling uniformly from the memory buffer the correlation between consecutive samples is reduced, leading to a more stable optimization of the action-parameter selection. 

\subsubsection{Exploitation vs. exploration}
Any RL algorithm using deterministic policy entails the trade-off between exploitation and exploration. For discrete action selection, we use the $\epsilon-greedy$ policy to ensure exploration, where with probability $\epsilon$ a random action is selected by sampling a uniform random distribution over all possible discrete actions. A high value for $\epsilon$ is set at the beginning to encourage exploration, but its value is reduced gradually over time until it reaches a certain minimum $\epsilon _{min}$, where it remains stable. 
\par
Ensuring the exploration of all possible continuous parameters is not possible. We use the approach proposed in  \cite{fujimoto2018addressing}, where a clipped zero-mean Gaussian noise is constantly added to the parameter selection policy (see Eqn. (\ref{e:exploration_noise})). This approach is motivated by the assumption that similar parameters should have similar costs and thus, similar estimates; and the noise addition is used to encourage exploration. After the addition of noise the parameter values are clipped to the allowed range $[p_{min},p_{max}]$, as defined in Section \ref{ss:actions}.
\begin{align}\label{e:exploration_noise}
\mu '\left(s, a\right)&=\text{clip}\left(\mu_{\theta_2}\left(s,a\right)+ w, p_{min}, p_{max}\right),\\
w \sim \text{clip}& \left(\mathcal{N}\left( 0,\sigma^2\right),-c,c\right),\notag
\end{align}
where $\sigma \text{ and }c$ are hyperparameters. Similarly to  the $\epsilon$ parameter for the action selection, we gradually reduce the value of $c$, until it reaches a minimum value $c_{min}$.

\subsubsection{Architecture}

The DNN architectures for the action, action parameter, and critic networks are the same. For all the networks, the inputs are processed by three fully connected layers consisting of 128-64 units respectively. Each fully connected layer is followed by a rectified linear unit (ReLU) activation function with negative slope $10^{-2}$. The weights of the fully connected layers are initialized using Xavier \cite{glorot2010understanding} initialization with a standard deviation of $10^{-2}$. 
\par 
The input of the actor action network is the network state, and connected to its final inner product layer there are $K+1$ linear outputs corresponding to the discrete action selection ($K$ servers plus the cloud). For the actor parameter network, the last layer comprises an hyperbolic tangent activation function scaled by $p_{max}$ and $\eta_{max}$ with two outputs, corresponding to the CPU and memory values allocated to the discrete action selected, while its inputs are the state and the  selected action. Finally  the critic network gathers the state, action, and the action parameters, and a single output value is obtained, the estimate of $Q(s,a,p)$. We use ADAM optimizer for both the actor and the critic, with a learning rate of $l_r$.
\subsection{Agent Cost Function}\label{ss:agent reward}
In Section \ref{sec: system model}, and more precisely in Eqn. (\ref{eqn:total cost}), we defined the global network cost as the main metric this work aims to minimize. However, we do not directly use ($\ref{eqn:total cost}$) as the metric the agent optimizes,  as we found it to be too general to guide the agent in its initial learning steps towards finding resource  allocation policies that lead to good results. The goal of this subsection is to define the cost function $\Psi^{\left(t\right)}$ that we use to provide feedback to the agent regarding its actions.\par  

Individual actions taken by the agent have direct impact in the VNF performance of the selected instance and then a lighter contribution on the total network performance. Thus, in order to guide the agent to learn to allocate resources to different VNF instances, we use the VNF instance cost of (\ref{eqn:vnf cost}). However, the minimization goal of this work is the total network cost, hence, we need to include it in the global picture. To this end we define the cost as follows:
\begin{equation}\label{eqn:training}
\Psi^{\left(t\right)}= \frac{C_{k,j}^{\left(t\right)}+\beta C_{T}^{\left(t\right)}}{\Gamma_{max}},
\end{equation}
where $\Gamma_{max}$ is a hyperparameter that guarantees  $\Psi^{\left(t\right)} \in [-1,1] $, while $\beta$ determines  how much the agent accounts for the total network cost. Eqn. (\ref{eqn:training}) is what we use in the DNN training, while  Eqn. (\ref{eqn:total cost}) is the objective function which is the aim of this work.
\par  
Furthermore, during the training phase, the proposed RL approach needs to learn the physics of the environment, that is, at the beginning of the learning process the agent might try to add/subtract more CPU or memory to a server than the one that is available/possible. In order to teach the agent the environment's physical limitations, whenever the algorithm outputs an infeasible action we offload the user  to the cloud, and impose a cost of $\Psi^{\left(t\right)}=-1$.

\section{Numerical Results}\label{s:NR}

In this section we present numerical results obtained with the PAT method described in Section \ref{s:PAT}. We start by presenting the DRL benchmarks to test the proposed algorithm, followed by the experimental setup and the parameters used in the simulations.

\subsection{DRL benchmarks.}
In order to assess the quality of the proposed algorithm, we compare the PAT agent with other DRL benchmark algorithms. 
\begin{itemize}
\item Greedy:  For each new user in the system, the \textit{greedy} algorithm checks whether the new user's VNF is already deployed in one of the CU servers. If so, computes the CPU and memory that the server would need to allocate to that VNF such that the new and existing users can be served from that VNF instance, i.e.,the resulting VNF's QoS lies in the range between $QoS^{min}$ and $QoS^{max}$. If  server's available resources allow for the calculated VNF resizing, the VNF is resized, and the user is allocated to that server (vertical scaling). If not, another server is checked until resources for the new user can be assigned (horizontal scaling). If no server is able to allocate this new user, the new user is offloaded to the cloud. 
\item Cloud: This approach offload all the traffic to the cloud. 
\item \textbf{DRL benchmarks}: To overcome the problem of discrete and continuous action selection formulated in this work, we use two distinct state of the art DRL algorithms. For server selection, discrete action, we use DDQN, while for parameter selection we use the following algorithms:
\begin{enumerate}
\item DDPG \cite{silver2014deterministic} with an hyperbolic tangent activation function in the outer layer scaled by the maximum values of the CPU and memory, respectively.
\item A3C \cite{mnih2016asynchronous}, where the output of the DNNs provide the mean and variance values of the Gaussian distributions used to sample the values of the CPU and memory. The parameter $T$ of \cite{mnih2016asynchronous}, is chosen to be $128$.
\item DDQN \cite{mnih2015human}, where we discretize the CPU and memory action, with a resolution of 5, meaning that the total number of actions is given by $\eta_{max}/5\times \rho_{max}/5$. 
\end{enumerate}
The DDQN for discrete action selection, and the previous set of algorithms are trained recursively (discrete action network training first, followed by parameter network training) for 1000 times. Each algorithm interacting with the environment a 10000 time-slots.
\end{itemize} 
\subsection{Parameters}
In our experimental setup, we consider $N=10$ VNFs that can be instantiated by the users to be deployed in the network, with features shown in Table \ref{t:VNFs}. We consider a CU with $K=10$ servers each with a CPU capability of $50 C\ \mathrm{Hz}$ ($\rho_{max}=50$) and memory capacity of $50 B$ bits ($\eta_{max}=50$).  The arrival rates ($\lambda_{j}^{\left(t\right) }$) for different VNFs at each epoch are sampled from a normal distribution, where each of the VNFs is characterized by different mean and variance values, listed in Table \ref{t:VNFs}. The values of the other parameters involved in the calculation of the cost function and those used to reinforce the actor behaviours are given in Table \ref{t:params}. The PAT algorithm parameters are collected in Table \ref{t:PAT}. We note that the values presented in Tables I, II and III for the numerical simulations are chosen as reasonable values that would lead to a solution with a balanced allocation of available resources in the servers and the cloud. Naturally, the value of these parameters in practice depends highly on the implementation and the technology used (memory/CPU capability) as well as the VNFs being considered; however, our problem formulation is general, and we have reached similar observations with a large variety of parameter values considered.
\vspace{1em}
\begin{table}[t]
\begin{center}
\captionof{table}{VNF resource requirements.} \label{t:VNFs} 
\begin{tabular}{ c|c|c|c|c|c|c|c|c|c|c |c} 
\toprule[1.5pt]
\multirow{2}{*}{\textbf{VNF}} &
\multicolumn{3}{ |c| }{\textbf{CPU}} &
\multicolumn{3}{ |c| }{\textbf{Memory}} &
\multicolumn{2}{ |c| }{\textbf{QoS}} &
\multicolumn{1}{ |c| }{\textbf{Others}} & 
\multicolumn{1}{ |c| }{\textbf{Mean}} & 
\multicolumn{1}{ |c }{\textbf{variance}} 
\\
\cline{2-12}
& $c_0$ & $c_r$ & $c_d$ & $s_0$ & $s_r$ & $s_d$ & $QoS_{min}$ & $QoS_{max}$ & $\gamma_j$ & $\mu_j$ & $\sigma_j$\\
\hline
$N_1$ & 3 & 5 & 4 & 6 & 5 & 3 & 35 & 70 & 2  &  2 &  1.5 \\
\hline
$N_2$ & 2 & 3 & 2 & 4 & 4 & 2 & 36 & 80 & 2  &  2.5 &  0.2 \\
\hline
$N_3$ & 1 & 4 & 2 & 2 & 3 & 2 & 27 & 63 & 2 &    4 &  0.5 \\
\hline
$N_4$ & 1 & 4 & 3 & 1 & 3 & 1 & 40 & 90 & 2 &   1 &  1 \\
\hline
$N_5$ & 2 & 6 & 2 & 3 & 4 & 3 & 20 & 100 & 2 &   2.5 &  1 \\
\hline
$N_6$ & 1 & 2 & 1 & 0 & 3 & 2 & 5 & 30 & 2 &  2 &  1.5 \\
\hline
$N_7$ & 2 & 3 & 2 & 2 & 5 & 3 & 56 & 80 & 2  &  5 &  1 \\
\hline
$N_8$ & 3 & 4 & 2 & 3 & 6 & 5 & 20 & 53 & 2 &   2 & 1 \\
\hline
$N_9$ & 1 & 4 & 3 & 4 & 4 & 2 & 40 & 90 & 2 &    3 &  0.5 \\
\hline
$N_{10}$ & 2 & 6 & 2 & 3 & 4 & 3 & 20 & 100 & 2 &  2 &  1 \\

\bottomrule[1.5pt]

\end{tabular}
\end{center}
\vspace{1em}
\end{table}

\vspace{0.5em}
\begin{table}[t]
\begin{center}
\captionof{table}{Delay parameters.} \label{t:params} 
\begin{tabular}{ c|c|c|c|c|c|c } 
\toprule[1.5pt]
\textbf{Delay ($ms$)} & $\delta _{r,c} = 3$ & $\delta _{r,s} = 4$ & $\delta _{d,b} = 20$ & $\delta _{d,t} = 10$ & \multicolumn{2}{|c}{ }   \\
\hline
\textbf{Cost ($m\$$)} & $ C_{r,s} = 3$ & $C_{r,p} = 6$ & $C_{i,o} = 2$ & $C_{i,v} = 1$ & $C_{c,0} = 1 $  & $C_{c,v} = 3 $\\
\bottomrule[1.5pt]

\end{tabular}
\end{center}
\vspace{1em}
\end{table}

\subsection{PAT performance}
The proposed PAT algorithm is run using 10 different seeds, and the average learning curves are depicted in Figure \ref{fig:learning}. Given the cost function in (\ref{eqn:total cost}), it can be seen that the agent maximizes the SLA to the point where the cost function is negative, meaning that the perceived QoS is greater than the weighted combination of the latency and the economical costs. Thus, given the predefined cost function, the agent learns to minimize the cost and to utilize the servers in an online manner. That is, since the agent generates its own dataset on the fly, by interacting with the environment, variations in the environment are directly fedback into the model by adding new traces into the memory buffer. Therefore, when a statistically significant change occurs, this is captured by the model.

\begin{table}[ht]
\begin{minipage}[b]{0.5\linewidth}
\centering
\begin{tabular}{  c || c   } 
\toprule[1.5pt]
Discount factor $\gamma$ = $0.99$ &
Target update $\tau $ = $5\cdot 10^{-3}$\\
$\epsilon$ = $0.8$ &
Learning rate $l_r $ = $10^{-3}$\\
$\epsilon_{min}$ = $0.05$ & 
Policy noise $\sigma$ = $0.2$ \\
$\epsilon_{decay}$ = $10^{-3}$ &
$c$ = $0.5$\\
$\left(\omega _1, \omega _2, \omega _3\right)$= $\left(1,1,2\right)$&  $c_{min}$ = $0.1$\\
$R_{min}=1$ & $\beta =0.2$\\
$t_{max}=100$ & $\Gamma_{max}=100$\\
\bottomrule[1.5pt]
\end{tabular}
\captionof{table}{PAT parameters.} \label{t:PAT} 
\end{minipage}
\begin{minipage}[b]{0.45\linewidth}
\centering
\includegraphics[width=1\textwidth]{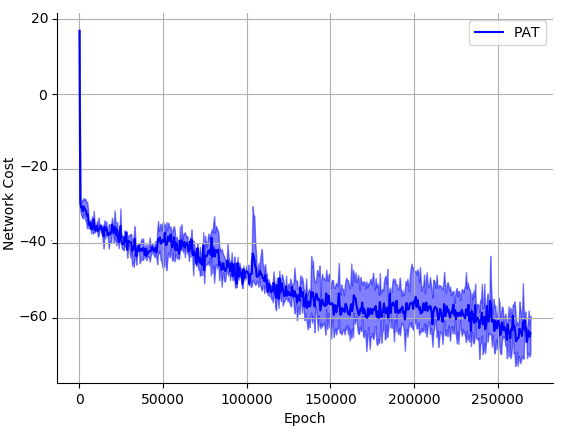}
\captionof{figure}{Evolution of the network cost in Eqn. (\ref{eqn:total cost}).}
\label{fig:learning}
\end{minipage}
\end{table}
\par 

\subsection{Mapping to KPIs}
We now define three KPIs for MANO in 5G networks, and map the results obtained with the PAT algorithm to these KPIs. The following KPIs are of interest for future 5G networks \cite{elasticity}:
\begin{itemize}
\item \underline{Resource utilisation  efficiency}: Given the CPU and memory resources, resource utilisation efficiency is defined as the ratio of utilized resources with respect the total available resources for the execution of a VNF, for a particular number of users. With the elastic functions employed in our model, the system should achieve a higher resource utilisation efficiency,  since  it  can  shelter a  much larger number of users over the same physical infrastructure.
\item \underline{Cost efficiency gain}: This metric captures the average cost of deploying and maintaining the network infrastructure to provide the required service to its users.  Given the elastic nature of the VNFs deployed, the CU should  be  able  to   optimally  dimension the network  such  that  less resources are required to support the same services; in addition,  the elastic system  should  avoid  the  usage  of  unnecessary  resources.
\end{itemize}
\subsection{PAT evaluation}
\begin{figure}[!h] 
\centering
\begin{subfigure}[b]{.5\textwidth}
  \centering
  \includegraphics[width=1\linewidth]{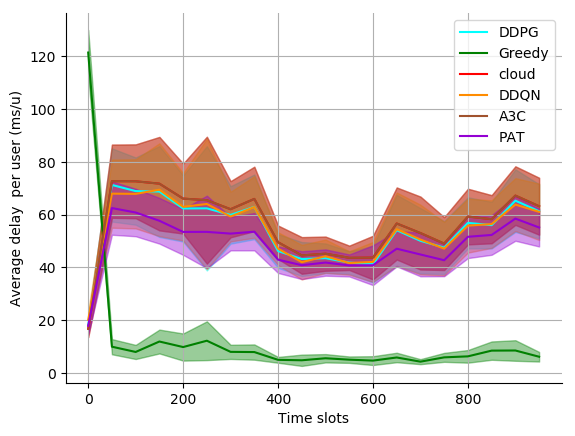}
  \caption{Latency cost (Eqn. (\ref{eqn:latency_cost})).}
  \label{sfig:delay}
\end{subfigure}
\begin{subfigure}[b]{.5\textwidth}
  \centering
  \includegraphics[width=1\linewidth]{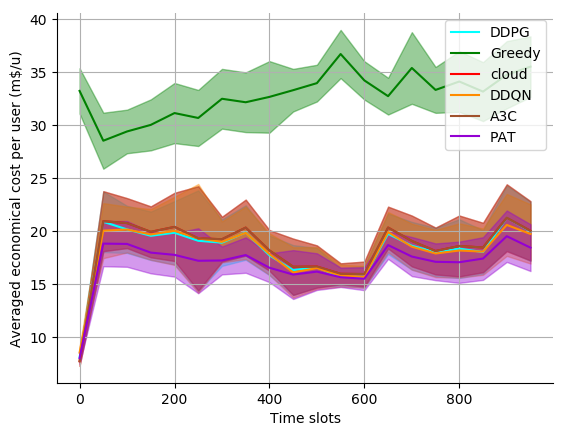}
  \caption{Financial cost (Eqn. (\ref{eqn:financial_cost})).}
  \label{sfig:economic}
\end{subfigure}
\begin{subfigure}[b]{.5\textwidth}
  \centering
  \includegraphics[width=1\linewidth]{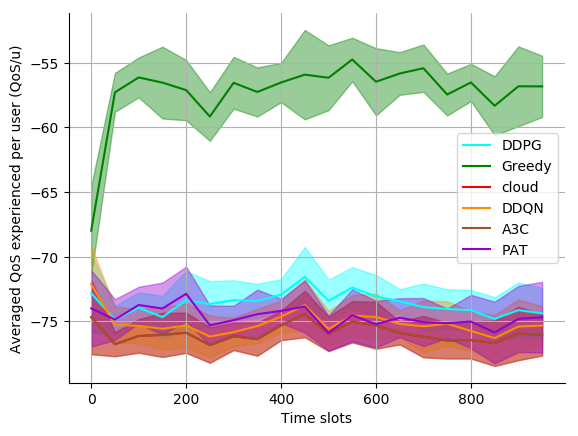}
  \caption{SLA cost (Eqn. (\ref{eqn:SLA})).}
  \label{sfig:qos}
\end{subfigure}
\begin{subfigure}[b]{.5\textwidth}
  \centering
  \includegraphics[width=1\linewidth]{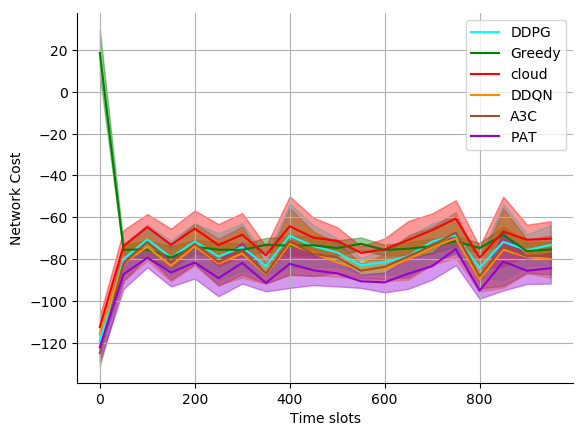}
  \caption{Network cost (Eqn. (\ref{eqn:total cost})).}
  \label{sfig:network cost}
\end{subfigure}
\caption{Defined costs comparison between PAT and the other DRL benchmarks. The shaded regions demonstrate the standard deviation of the average evaluation over 10 trials. \label{fig:bench}}
\end{figure}

In Figure \ref{fig:bench} and \ref{fig:resources}, we present the results of the proposed PAT approach and that of the other benchmark algorithms. The comparison is carried out under exactly the same traffic patterns, i.e., same arrival and departure times.
\par 
From Figures \ref{sfig:delay}, \ref{sfig:economic} and \ref{sfig:cloud} it can be observed that the particular set of parameters chosen for this network configuration entails that users offloaded to the cloud experience higher delays than the users served by the CU. On the contrary, for the financial cost, VNFs instantiated in the CU have a higher cost than VNFs instantiated in the cloud. That explains why the scheme with lowest cloud utilization, i.e., greedy, has the lowest delay per user but the highest financial cost per user. \\
Furthermore, Figure \ref{sfig:cloud} shows how all the  DRL algorithms decide to allocate more users to the cloud than the CU, this is mainly for two reasons affecting the learning process:
\begin{enumerate}
    \item \underline{The VNF cloud allocation does not carry any penalty}. Contrary to VNF allocation in the CU, where algorithms are penalised if the physical limitations of the servers are not respected or allocated resources are not enough to fulfill the SLA, the deployment of VNFs at the CU carry a positive reward.
    \item \underline{The QoS drives the learning experience.} Since $\omega_3$ is greater than the other two weights scaling the costs functions defined in Eqn. (\ref{eqn:total cost}) and (\ref{eqn:vnf cost}), algorithms aim to maximize SLA cost (given the RL reward function of (\ref{eqn:training})). Thus, as allocating VNF to CU may lead to lower QoS if the CU fails to provide the maximum demanded resources, the DRL tends to use the cloud where $QoS^{max}$ is guaranteed.  
\end{enumerate}

\subsubsection{Resource utilisation  efficiency}
\begin{figure}[!h] 
\centering

\begin{subfigure}{.5\textwidth}
  \centering
  \includegraphics[width=1\linewidth]{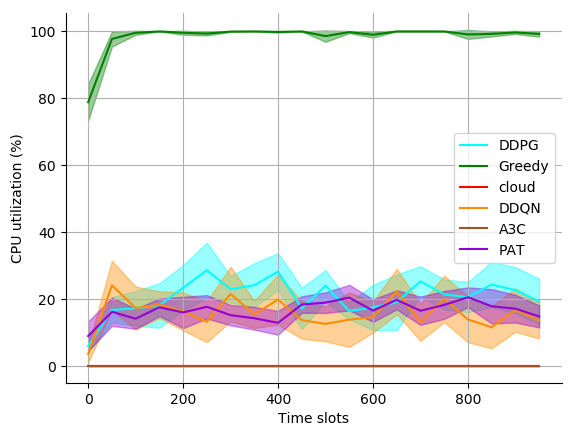}
  \caption{CPU utilization.}
  \label{sfig:CPU}
\end{subfigure}%
\begin{subfigure}{.5\textwidth}
  \centering
  \includegraphics[width=1\linewidth]{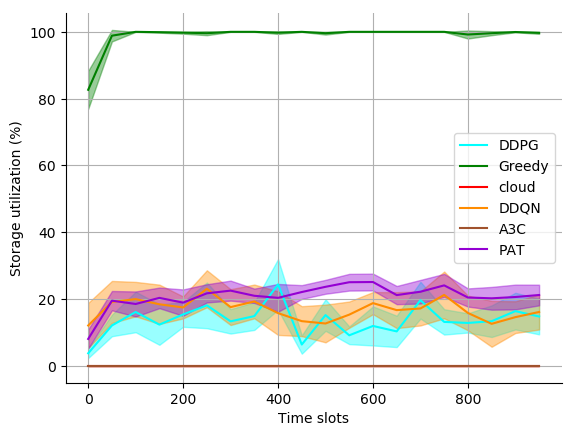}
  \caption{Memory utilization.}
  \label{sfig:memory}
\end{subfigure}
\begin{subfigure}{.5\textwidth}
\centering
  \includegraphics[width=1\linewidth]{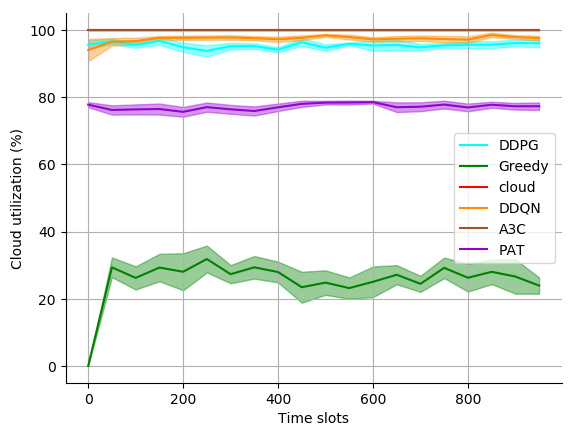}
  \caption{Cloud utilization.}
  \label{sfig:cloud}
 \end{subfigure}
\caption{Resource utilization.  \label{fig:resources}}
\end{figure}

  A comparison of the resource utilisation is presented in Figure \ref{fig:resources}. The figure shows that the PAT algorithm leads to a more efficient usage of the CPU and memory resources compared to A3C, DDPG and DDQN, as for a similar CPU and memory utilization (Figures \ref{sfig:CPU} and \ref{sfig:memory}) a lower amount of traffic is offloaded to the cloud (Figure \ref{sfig:cloud}). The efficient usage of resources accomplished by PAT is also visible in Figures \ref{sfig:delay} and \ref{sfig:economic}, where even though similar resources are being utilized by the the other DRL algorithms, the latency cost and the financial cost achieved by the PAT are lower (on average). The greedy approach aims to allocate as many users as possible to the CU, that is why its CU resources are fully utilised most of the time and the average delay of the users is the lowest, while its financial cost is the highest.

\par 
\subsubsection{Cost efficiency gain} 
The comparison between the averaged economical cost of the PAT deployment and the other schemes is presented in Figure \ref{sfig:economic}, where a gain in economical cost by the PAT algorithm is clearly visible. The economical cost difference between the PAT and the greedy is straightforward, as the network configuration used entails a higher cost for the use of CU resources. The reduction of financial cost of the PAT compared to the other DRL algorithms is because the latter allocate more CPU and memory than needed, given the fact that most of their traffic is directed to the cloud, such that the CU resources are underutilized, incurring higher cost per user.

\subsubsection{Network Cost}

Figure \ref{sfig:network cost} shows that the proposed PAT algorithm outperforms the other approaches on the main metric of this work, the network cost defined in Eqn. \ref{eqn:total cost}. The PAT finds a middle point, between directing traffic to the CU and offloading it to the cloud. One of the reason that the proposed approach outperforms DRL benchmarks is that the optimization of the server selection and the resource allocation is done jointly. Contrary to A3C, DDQN and DDPG, where the training is done by iterations, the PAT algorithm propagates the gradient of the value-estimates obtained by the critics to the parameter network and the server selection network at the same time, pushing both networks to more optimal points simultaneously, improving the efficiency of each training update. 
\par
\begin{figure}
\centering
  \includegraphics[width=0.6\linewidth]{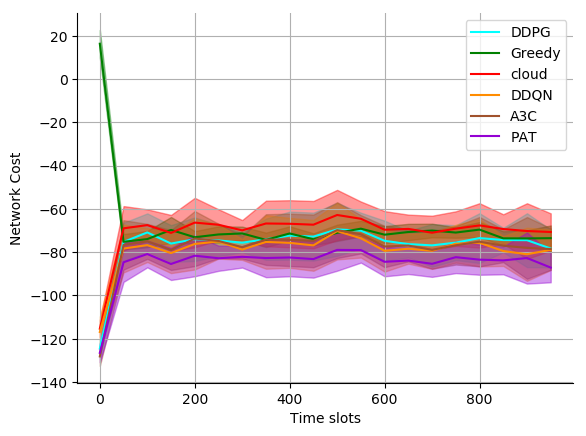}
  \caption{Network cost with constants $\lambda_j$.}
  \label{fig:constant_statistics}
\end{figure}

It is somewhat surprising that the greedy algorithm performing comparably, or sometimes even better than the baseline DRL algorithms at some time periods. This is because the DRL algorithms, in contrast to PAT, have delays in adapting to the randomness of the environment, or tend to slightly over-provision resources to be able to cope with highly time-variant traffic demands. If, however, we keep the traffic statistics (arrival rates) constant, we can see in Figure \ref{fig:constant_statistics} that baseline DRL algorithms outperform greedy, although PAT is still the best performing algorithm. This shows that PAT not only outperforms other baselines in exploiting the resources in the most efficient manner in a static environment, but also is the fastest in terms of adapting to variations in the environment.

\section{Conclusions and Future Work}\label{s:conclusions}
We presented a novel DRL algorithm for autonomous management and orchestration of VNFs, where the CU learns to re-configure resources (CPU and memory), to deploy new VNFs instances, and to offload VNFs to a central cloud depending on network conditions, VNFs' requirements and the pool of available resources. We formulated the stochastic problem of resource allocation for a radio access network, where a group of BSs  are connected to a CU that needs to provision and manage resources for the BS users as a Markov decision problem. Then, we proposed a DRL-based solution for this MANO problem, more precisely, we presented a novel approach named PAT, which leverages the actor-critic method to learn to provision network resources to the VNFs in an online manner, given the current network state and the requirements of the deployed VNFs. The novel architecture implements two critics for action value function estimation (twin), and two actor networks are used to determine the action and the parameter. A deterministic policy is implemented for both action and parameter selection. We have shown that the proposed solution outperforms all benchmark DRL schemes as well as heuristic greedy allocation in a variety of network scenarios, including static traffic arrivals as well as highly time-varying traffic settings. To the best of our knowledge, this is the first work that considers DRL for network MANO of VNFs.
\par 
As future research directions, we consider addressing the MANO of VNF chains. The problem addressed in this work does not take into account the likely relation between different VNFs to form VNF chains, where NFs may have a temporal ordering in which they are requested by users. This factor highly increases the complexity of resource allocation as the overall user experience might be affected by a subtle VNF resource modification. Furthermore, another interesting direction might be the use of hierarchical DRL approach to be able to coordinate at the same time the network slice requirements with the sub-VNFs instantiated by those slices when deployed in a shared pool of resources.
\bibliographystyle{IEEEtran}
\begin{center}
\bibliography{slicing}
\end{center}

\end{document}